\documentstyle[11pt,newpasp,twoside,epsf]{article}
\def\edcomment#1{\iffalse\marginpar{\raggedright\sl#1\/}\else\relax\fi}
\reversemarginpar

\begin{document}
\title{A Solar Neighborhood Search for Tidal Debris from $\omega$ 
Centauri's Hypothetical Parent Galaxy}
\author{Dana I. Dinescu\altaffilmark{1}}
\affil{Department of Astronomy,
University of Virginia, 530 McCormick Road, Charlottesville, VA 22903}
\altaffiltext{1}{Astronomical Institute of the Romanian Academy, Str.
Cutitul de Argint 5, RO-75212, Bucharest 28, Romania}

\begin{abstract}
Recent stellar population and chemical abundance studies point to
an accreted origin of $\omega$ Cen. In this light, and given the 
retrograde, small size orbit of $\omega$ Cen, we search for a kinematical
signature left by its hypothetical parent galaxy in the Solar
neighborhood. We analyze the largest-to-date sample of metal poor stars
(Beers {\it et al.} 2000) and we find that, in the metallicity range
$-2.0 <$ [Fe/H] $\le -1.5$, a retrograde signature that departs 
from the characteristics of the inner halo, and that resembles
$\omega$ Cen's orbit, can be identified.
\end{abstract}

\section{Introduction}
Recent advances in understanding the nature and origin of 
the highly unusual globular cluster $\omega$ Centauri 
 (see Majewski  {\it et al.} 2000
for a summary of properties), are due primarily 
to the following findings:  the multiple-peak metallicity 
distribution seen in the structure of the giant branch 
(Lee {\it et al.} 1999; Pancino {\it et al.} 2000; 
Frinchaboy {\it et al.} 2001), 
the correlation between age and metallicity (e. g., Hughes \& Wallerstein 2000,
Hilker \& Richtler 2000), 
and the s-process enhanced enrichment in cluster stars compared
to halo stars of similar metallicity (Smith {\it et al.} 2000; Vanture,
Wallerstein \& Brown 1994).
These findings suggest that $\omega$ Cen underwent self-enrichment
with at least three primary enrichment peaks
(Pancino {\it et al.} 2000, Frinchaboy {\it et al.} 2001), over 
a period of at least 3 Gyr (Hughes \& Wallerstein 2000). The 
s-process heavy-elements are primarily 
synthesized in low-mass (1.5 to 3.0 M$_{\odot}$)
asymptotic giant branch (AGB) stars (see e.g., Travaglio {\it et al.} 1999
and references therein). 
In order to enrich the cluster in 
s-process elements, the ejecta from low-mass stars that evolve on
timescales of $10^{9}$ years had to be retained by the cluster and
incorporated in the next generations of stars. This long and complex
star formation history is inconsistent with the cluster originating
on its current orbit, which is of low energy and confined to the disk.
With a period of only 120 Myr (Dinescu, Girard, \& van Altena 1999 -
hereafter DGvA), the frequent disk crossings would
have certainly swept out all of the intracluster gas soon after its
formation, and the result would be a single-metallicity 
system that would resemble most of the Galactic globular clusters.

It appears thus that $\omega$ Cen evolved
somewhere away and independently from the Milky Way, in a system that was
massive enough to retain ejecta from previous generations
of stars, and to undergo multiple episodes of star formation.
Its current orbit can be reconciled with the complex star formation
history only if it represents a strongly decayed orbit.
This, in turn, requires a massive enough system such that dynamical friction 
was able to drag it to the inner regions of the Galaxy.
This system must have also been rather dense in order to
survive the tidal field of the Milky Way and continue
to loose orbital energy due to dynamical friction down to an orbit with an
apocenter of the order of the Solar circle radius.
The current mass of $\omega$ Cen (5 $10^{6}$ M$_{\odot}$; Meylan {\it
et al.} 1995) can not generate sufficient dynamical friction
to modify its orbit to its current small size (DGvA). 

Following these arguments, the debris from the massive putative parent
galaxy of $\omega$ Cen may be expected to imprint a kinematical feature in
large samples of local, metal-poor stars. 
The purpose of this investigation is to search for such a feature in the
kinematically hot halo. 
We have used three data sets: the largest, kinematically unbiased
sample of metal poor stars ($\sim 1200$) provided by 
Beers {\it et al.} (2000) (hereafter B2000),
the sample of globular clusters with measured absolute proper
motions (DGvA updated with new distances from Harris 1996, and
with a few more clusters; see Dinescu {\it et al.} 2001), and a 
small sample of stars with complete kinematics and abundance
measurements for O, Na, Mg, Si, Ca, Ti, Cr, Fe, Ni, Y and Ba
(Nissen \& Schuster 1997, hereafter NS97).
\section{$\omega$ Cen and the Globular Cluster System}
DGvA pointed out that $\omega$ Cen's orbit is markedly
retrograde for its low inclination and low orbital energy, when compared
to the orbits of clusters with similar metallicity, horizontal
branch morphology and orbital energy (see Fig. 6 and 7 in DGvA). 
Along with $\omega$ Cen, two other clusters were identified to
lie in the same region of the orbital parameter space:
NGC 362, and NGC 6779 (M56) (DGvA). They both have low-inclination,
retrograde orbits, with pericentric distances similar to
that of $\omega$ Cen ($\sim 1$ kpc). However, NGC 362 and NGC 6779
have apocentric distances larger than $\omega$ Cen: 11 and 14 kpc
respectively, compared to 6 kpc for $\omega$ Cen. 

Some controversy regarding NGC 362's absolute proper motion is apparent:
Geffert reported at this conference that his new determination with 
respect to $Tycho$ stars implies an orbit less similar to that of 
$\omega$ Cen. In our paper (DGvA) we have used an average of two proper-motion
measurements for NGC 362. These measurements were:
a) calibrated to $Hipparcos$ stars (Odenkirchen {\it et al.} 1997),
and b) calibrated to stars in the Small Magellanic Cloud (SMC)
(Tucholke 1992), to which was applied the absolute proper motion of the
SMC as measured by Kroupa \& Bastian (1997). 
While the absolute proper motion determination may remain debatable, 
another piece of evidence was presented at this conference regarding 
a possible common origin of NGC 362 and $\omega$ Cen.
Smith {\it et al.} (2000) showed that 
the [Cu/Fe] abundance of $\omega$ Cen's stars is low 
([Cu/Fe] = -0.6) and remains remarkably
constant with increasing metallicity (their Fig. 9). 
In field stars, the Cu abundance increases
with metallicity (Sneden, Gratton, \& Crocker 1991, Castro, Porto de Mello,
\& da Silva 1999) from subsolar values to 0 at solar metallicity. This trend
is explained if a substantial contribution to Cu comes from supernovae Type Ia
(Matteucci et al. 1993). Cuhna {\it et al.} (2001) 
reported at this conference that NGC 362 
too shows the deficit in [Cu/Fe] much like $\omega$ Cen's stars, and 
in contradistinction with clusters NGC 288 and M4. These latest
clusters have a similar metallicity with NGC 362, but do
not show an underabundance in Cu, when compared to field stars of similar
metallicity.

NGC 6779's absolute proper motion determination was based only on the
calibration to $Hipparcos$ stars (Odenkirchen {\it et al.} 1997). 
This measurement seems reliable
as diffuse X-ray emission detected in the area of the cluster is
found to be aligned with the direction of the proper motion 
(Hopwood {\it et al.} 2000).
This emission is interpreted as heated interstellar medium in the
wake of the cluster as a result of the interaction of the intracluster gas
with the halo gas.
\section{Inner Halo Properties as Derived from the Beers {\it et al.}
Catalog}
Chiba \& Beers (2000) characterized the local halo from the analysis of
 the  B2000 catalog. The sample they analyzed comprised stars within
4 kpc of the Sun, and with Galactocentric distance along the plane
between 7 and 10 kpc. From their results we summarize here 
those that are relevant to our investigation. 
A relatively ``pure'' sample of halo stars
can be found at [Fe/H] $\le -2.0$ (their Fig. 9). Thick disk stars
begin to contribute to the overall metal poor population
([Fe/H] $< -1$), with 
a fraction that increases with increasing metallicity (their Fig. 10).
In the metallicity range $-2.0 <$ [Fe/H] $\le$ -1.5, this fraction is
$\sim 10 \%$. The mean rotation velocity of halo stars decreases
with increasing distance from the Galactic plane: $-52 \pm 6$ km/s/kpc). 
At a mean distance from the Galactic plane $|z| \sim 0.5$ kpc, 
the rotation velocity is $\sim 55$
km/s (their Fig. 4). They have also found that, at a metallicity of -1.7,
the mean rotation of the low ($|z| < 1$ kpc) halo drops to 20 km/s, from
60 km/s at lower metallicities. At higher metallicities, the mean rotation
increases with increasing metallicity, as more thick disk stars 
begin to contribute to the overall population (see their Fig. 3).
At the same value of [Fe/H] $\sim -1.7$, Chiba \& Beers (2000) also 
find a higher concentration of stars at large eccentricities (e $\sim 0.9$)
(their Fig. 6). They speculate that stars at [Fe/H] $\sim -1.7$ formed
from infalling gas rather than from the somewhat more organized material
of the inner halo. Interestingly, the mean metallicity of 
$\omega$ Cen is -1.6 (Harris 1996).
\begin{figure}
\plotone{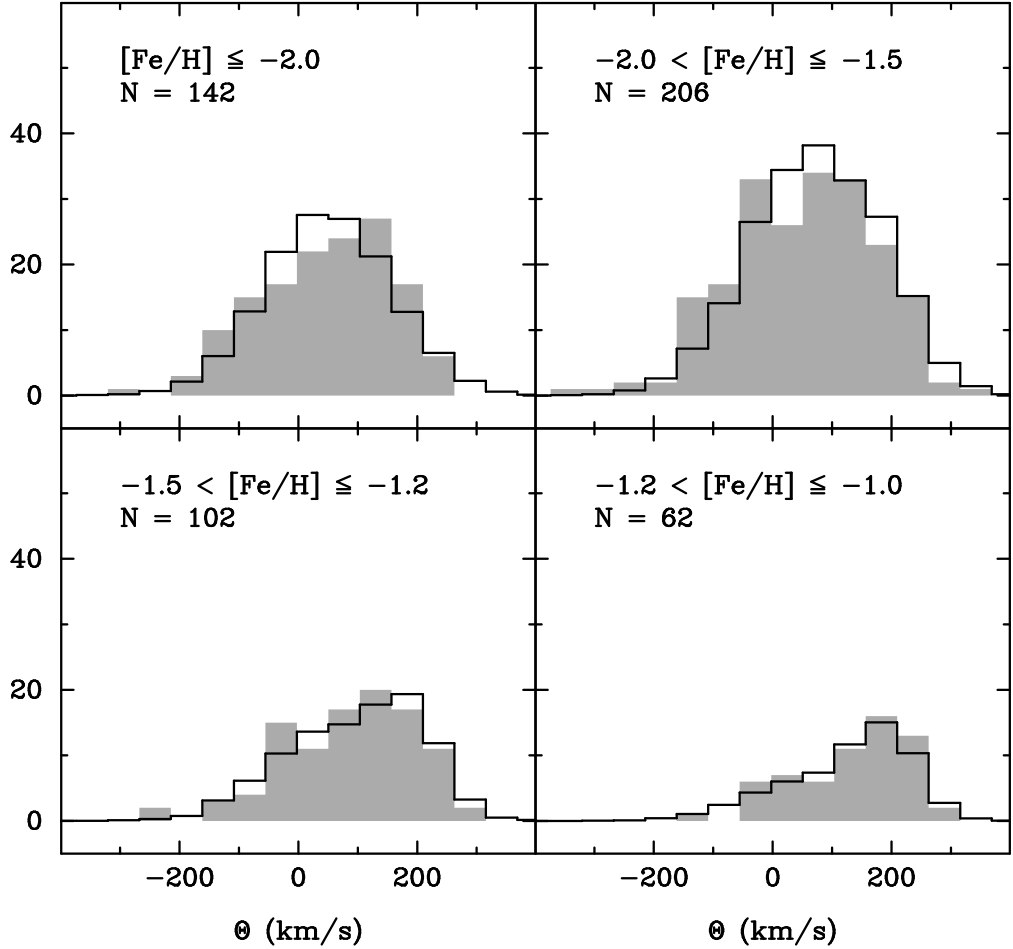}
\caption{Rotation velocity distributions for four metallicity intervals,
and for $|z| \le 1$ kpc.
The shaded areas represent the observed distributions, while the line shows
the distributions derived from a simple kinematical model (see text).}
\end{figure}

Here, we will look in more detail at the stars that produce the drop
in the rotation velocity dependence with metallicity. 
Using the B2000 distances, radial velocities, and absolute proper
motions we derive velocities using R$_0$ = 8 kpc, $\Theta_{LSR} = 220$
km/s , and a peculiar Solar motion (U, V, W ) = (-11.0, 14.0, -7.5) km/s,
where U is positive toward the Anticenter.
We integrate the orbits in the axi-symmetric analytic potential   
described in Paczy\'{n}ski (1990), 
and we derive orbital parameters in the manner described in DGvA.
\subsection{Rotation Velocity Distributions}
In Figure 1 we show the distribution of the rotation velocity 
$\Theta$ (in a cylindrical coordinate system, where the $\Pi$
component is positive outward from the Galactic center, 
$\Theta$ is positive toward Galactic rotation, and W is positive
toward the North Galactic Pole), for four metallicity groups, and
within $|z| \le 1$ kpc.
The shaded histograms represent the observed distribution, and the 
continuous line represents the distribution derived from a simple
kinematical model. Velocities in this model were drawn from Gaussian
velocity distributions of means and dispersions derived from 
the B2000 catalog, and including both halo and thick disk populations.
Specifically, for [Fe/H] $\le -2.0$ we used $<\Theta_{halo}> = 50$ km/s,
$\sigma_{halo} = 106$ km/s, and the fraction of thick disk stars
f$_{TD}$ = 0.0; for $-2.0 <$ [Fe/H] $ \le -1.5$,  $<\Theta_{halo}> = 60$ km/s,
$\sigma_{halo} = 106$ km/s, f$_{TD}$ = 0.1, $<\Theta_{TD}> = 190$ km/s,
$\sigma_{TD} = 50$ km/s; for  $-1.5 <$ [Fe/H] $ \le -1.2$,  $<\Theta_{halo}> = 60$ km/s,
$\sigma_{halo} = 106$ km/s, f$_{TD}$ = 0.3, $<\Theta_{TD}> = 190$ km/s,
$\sigma_{TD} = 50$ km/s; and for $-1.2 <$ [Fe/H] $ \le -1.0$,  $<\Theta_{halo}> = 60$ km/s,
$\sigma_{halo} = 106$ km/s, f$_{TD}$ = 0.5, $<\Theta_{TD}> = 190$ km/s,
$\sigma_{TD} = 50$ km/s. The total number of stars in each sample is
indicated in the respective panel.
The model distributions were drawn for 10000 points, and 
normalized to the area of the observed distributions.
We have chosen only one mean rotation velocity to represent the halo,
for each metallicity interval in the following way. For all the stars 
within $|z| \le 1$ kpc, the median $z$ was determined, and 
for this value of $z$, the rotation velocity was read directly
from the linear fit derived in Fig. 4 of Chiba \& Beers (2000). 

For the ``pure'' halo sample (top, left panel of Fig. 1) the simple,
one-Gaussian model does not represent the data well.  The data peak at 
$\Theta \sim 120$ km/s --- a higher velocity than the value we chose 
in the model --- and they also show more stars at strongly
retrograde velocities ($\Theta \sim -200$ km/s) than the model.
A superposition of two or more Gaussians with appropriately 
chosen mean rotation velocities and relative number of stars 
with respect to the
majority of stars that peak at $\Theta = 120$ km/s, 
can be envisaged however, to better describe the data.
This composite kinematics can be thought of as arising
from the fact that some of the stars within $|z| \le 1$ kpc
are visiting this region rather than being confined to it. These stars  
actually reside on orbits of various, mean $z$ larger than 1 kpc, 
and therefore belong to various populations of accordingly
lower mean $\Theta$.

The next two metallicity samples (upper right and lower left panels)
 show the expected peaks at 
prograde velocities that are approximately matched by the
models, and a second peak at modest retrograde velocities 
($\Theta \sim -30$ km/s). 
The most metal rich sample (lower right panel) shows a rather good fit to the
simple kinematical model, but there are few stars in this sample. 
The second peak seen at retrograde velocities
is intriguing not necessarily because it is not reproduced
by the kinematical model, but for the following reason. The sample
with $-2.0 <$ [Fe/H] $ \le -1.5$ has a minor contamination from the thick 
disk; therefore, it can be compared in a meaningful way to the 
``pure'' halo sample. They both show a tail toward strongly retrograde
velocities --- that is not reproduced by our one-Gaussian models --- but
there is no second peak in the ``pure'' halo sample. 
It is thus difficult to explain the second peak at retrograde velocities
even in a composite kinematical model of the halo, as sketched above and
perhaps seen in the ``pure'' halo sample. One would have 
to imagine a significant population that comes from orbits
of one particular mean $z$ value that corresponds to  $\Theta = -30$ km/s,
rather than a population drawn from a mixture of various mean $z$ orbits.

In what follows we will focus on two metallicity samples:
the ``pure'' halo [Fe/H] $\le -2.0$, and the sample within $-2.0 <$ [Fe/H] $ \le -1.5$, which has a relatively small fraction of thick disk stars
($10\%$).

\subsection{Orbital Angular Momentum Distributions} 

We have constructed the orbital angular momentum L$_{z}$ distributions
for the [Fe/H] $\le -2.0$ sample and for the $-2.0 <$ [Fe/H] $ \le -1.5$
sample, and for stars with maximum excursions above/below the Galactic plane
$|z_{max}| \le 4$ kpc, to represent the low halo, and $|z_{max}| \le 20$ kpc
for the whole halo. These are shown in Figure 2, where the thin line
represents the metal-poor sample (referred to as sample A hereafter), 
and the thick line the less-metal-poor 
sample (referred to as sample B hereafter); 
the range of $z_{max}$ is specified in each panel. 
\begin{figure}
\plotone{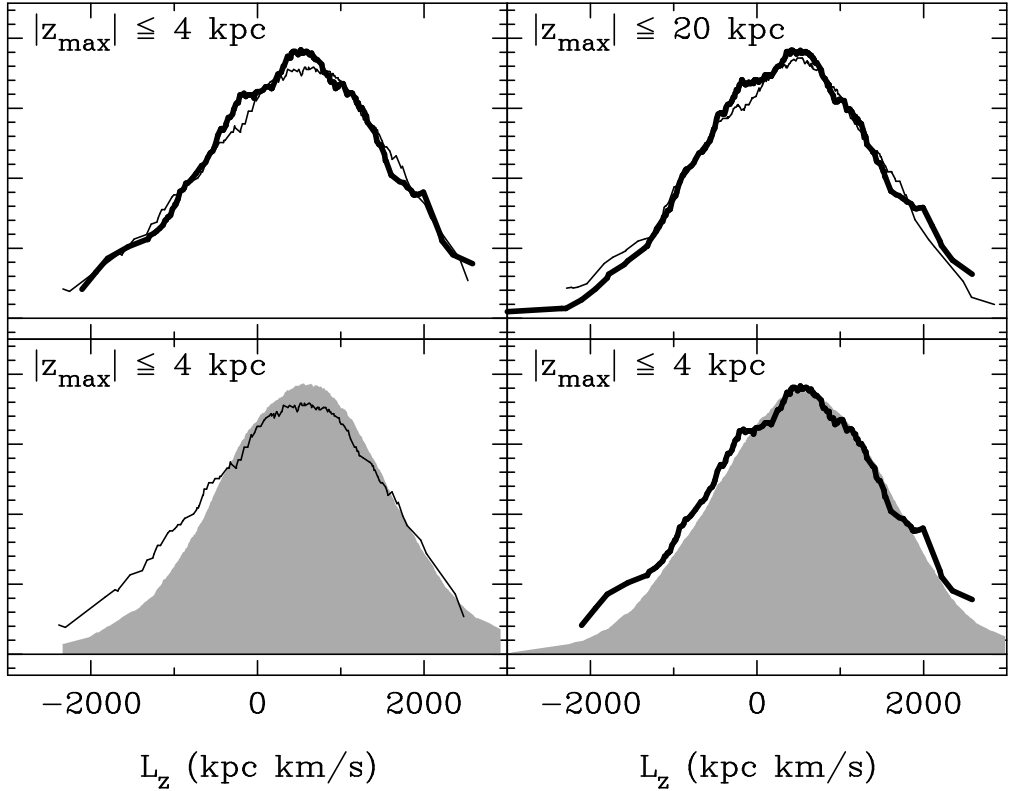}
\caption{Orbital angular momentum distributions. The thin line represents
the sample with [Fe/H] $\le -2.0$ (sample A), while the thick line that with 
$-2.0 < $ [Fe/H] $\le -1.5$ (sample B). The shaded areas represent the distributions as derived from a kinematical model (see text).
The top left panel, and the bottom panels show stars that are more confined to the Galactic plane than the stars in the top right panel.}
\end{figure}
The distributions
were constructed by passing a moving box of half-width equal to the
dispersion in L$_{z}$: 900 kpc km/s. They were also normalized to the
total number of stars, in order to be intercompared. A thick-disk component
of $\sim 10 \%$ (Section 3) of the total number of stars
is expected in sample B, roughly at L$_{z} = 8$ kpc $\times 190$ km/s = 1520 
kpc km/s. For this reason, sample B was normalized to a slightly larger number 
than the number corresponding to a ``pure'' halo population.  Therefore,
the distributions were slightly shifted in L$_{z}$ (up to
100 kpc km/s) in an attempt to best match them, and to keep
 the sample B distribution slightly under the sample A distribution.
For the $|z_{max}| \le 20$ kpc samples (top right panel), the 
distributions are remarkably similar with the following exceptions. 
They differ in the thick-disk range (L$_{z} = 1500$ to 2000 kpc km/s), 
which was expected, at strongly
retrograde orbits (L$_{z} \sim  -1800$ kpc km/s), but there are few stars here,
 and --- most significantly --- at modest retrograde orbits L$_{z} \sim -400$
kpc km/s. At this latter value of L$_{z}$, sample B shows an excess of stars,
that produce a shoulder in the distribution when compared with sample A.
For the low halo ($|z_{max}| \le 4$ kpc; top left panel in Fig. 2), sample B
again shows the shoulder, or the excess of stars at L$_{z} \sim -400$, 
when compared to sample A. Sample B also seems to have a slightly 
sharper peak than sample A. 

The bottom panels of Fig. 2 show --- for the low halo --- the observed
distributions for sample A (left panel) and sample B (right panel) compared
to a model of the halo (shaded curves).
This model was designed to describe the halo only,
and to input the kinematics only and not the spatial distribution.
Thus, for each star in the appropriate metallicity sample, a velocity
drawn from a Gaussian distribution was assigned, while preserving
the positional information from the B2000 catalog.
Each of the velocity components were
drawn from Gaussian distributions specific for the halo.
We have used ($<\Pi>, <\Theta>, <$W$>$) = (0, 60, 0) km/s, and
($\sigma_{\Pi}, \sigma_{\Theta}, \sigma_{W}$) = (141, 106, 94) km/s.  
Ten sets of halo velocity components were assigned to each star, 
and for each the orbits were integrated in order
to obtain the model distributions. The models agree well with the
data in the prograde regime; however, they fail to reproduce the tail
toward retrograde orbits, likely due to the use of a
single value of the mean $\Theta$. They can not reproduce the shoulder
structure seen in sample B. 
Therefore, preserving the current spatial distribution 
of the stars and statistically assigning halo-like velocities will not
 produce the excess of stars at L$_{z} \sim -400$ kpc km/s.
The conclusion is that the shoulder 
is due to stars on particular orbits; the excess of such orbits
is not seen in sample A. For Solar neighborhood stars, this L$_{z}$ 
excess corresponds to that seen in the rotation velocity distribution
at modest retrograde velocities (Section 3.1).

\section{Orbit Characteristics}

We look now at the detailed characteristics of the orbits in samples A and B.
In Figure 3, top panels, we show the orbital energy E as a function 
of L$_{z}$. All left panels represent sample A, while the right panels
sample B. The number of stars is indicated in each panel. Open symbols
represent all stars in B2000, while highlighted (filled)
 symbols show those classified
as RR Lyrae variables in B2000. For sample B, an almost vertical
structure can be seen at L$_{z} \sim -400$ kpc km/s and orbital energy 
$-1.2~10^5 \le$ E $< -0.5~10^5$ (km/s)$^2$. This structure is seen in the
RR Lyrae population as well.
\begin{figure}
\plotone{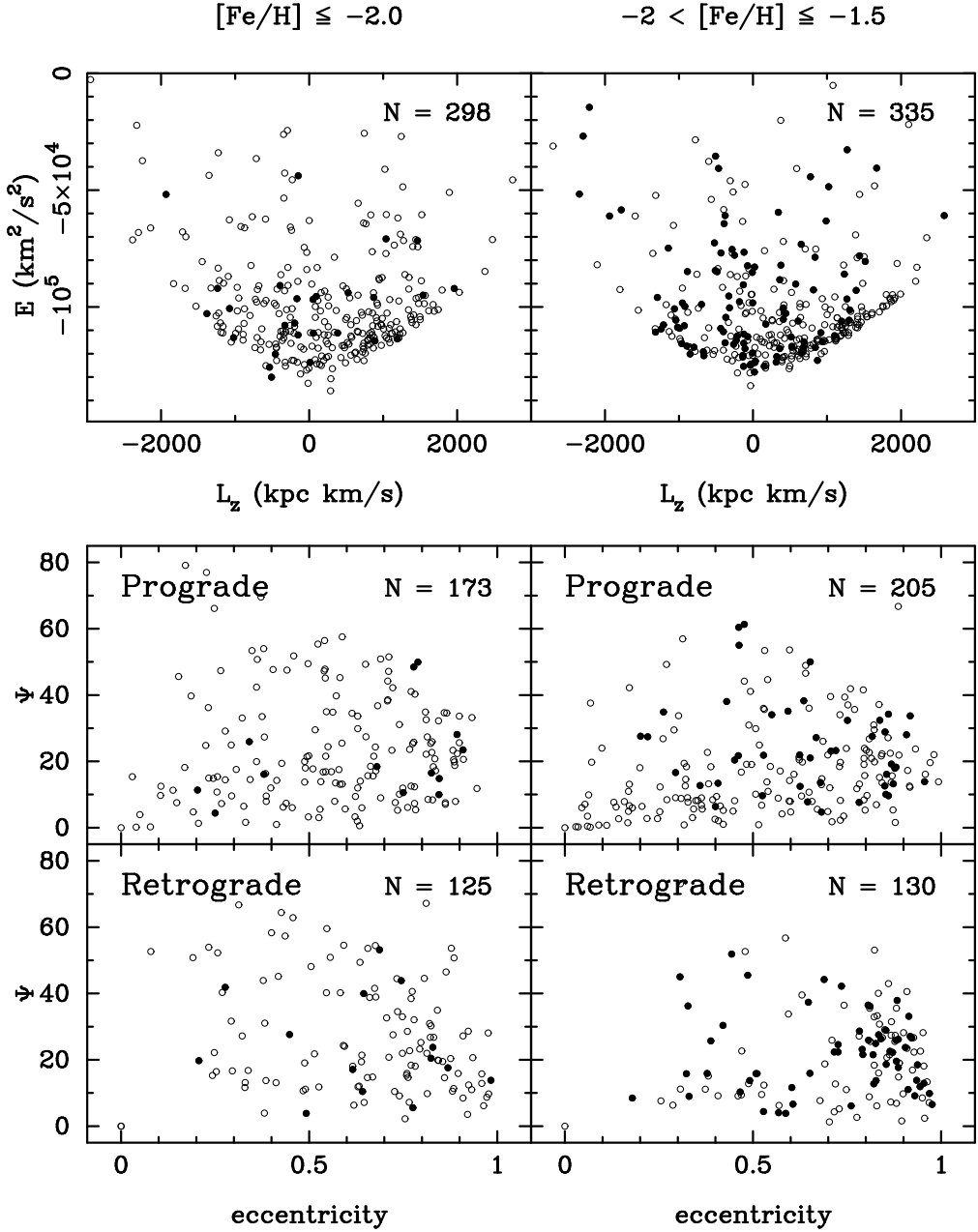}
\caption{Orbital parameters for Beers {\it et al.} (2000) stars. All left side
panels show stars with [Fe/H] $\le -2.0$ (sample A), while the right side ones show those with $-2.0 < $ [Fe/H] $\le -1.5$ (sample B). 
The filled circles represent the RR Lyrae
variables in the B2000 catalog. The top panels show the orbital energy E
as a function of orbital angular momentum. The rest of the panels show 
the orbital inclination as a function of eccentricity for prograde orbits
(L$_{z} > 0$; middle panels), and for retrograde orbits (L$_{z} \le 0$;
 bottom panels).}
\end{figure}
We divide now each metallicity sample into two groups: prograde
(L$_{z} > 0$) and retrograde (L$_{z} \le 0$) orbits. For each group
we plot the orbit inclination $\Psi$ as a function of orbital eccentricity $e$.
The middle panels of Fig. 3 show the prograde group, 
and the bottom panels the retrograde group. 
A higher density of stars, reproduced as well in the RR Lyrae population,
can be seen clumping at $e \sim 0.85$ and $\Psi \sim 25\deg$, for the
retrograde group of the B sample of stars.

By selecting stars with high eccentricities, we safely discard the 
poorly known fraction of stars that have rotational support,
be it halo or thick disk. Therefore, in each metallicity sample, one 
would expect the same number of highly eccentric stars in the prograde 
sample, as in the retrograde sample, in a completely pressure-supported
halo. Selecting stars with $e > 0.8$ we find
33 stars in the prograde sample A, and 37 in the retrograde
sample A, while in sample B we find 45 stars in the prograde group and 70
in the retrograde group. The 2.3-$~\sigma$ excess of
highly eccentric stars in the retrograde group  of sample B
resides at moderate to very low orbit inclinations.
Interestingly enough, the RR Lyrae population in sample B follows
the same pattern at $e > 0.8$: there are 15 stars in the prograde group, 
and 30 stars in the retrograde group.

We plot now all of the three integrals of motion, L$_{z}$, E, and 
total angular momentum L in Figure 4. 
In addition to the B2000 sample, we use the NS97 sample of stars
with chemical abundance measurements, and the globular cluster sample
(DGvA). Distances and absolute proper motions for the NS97 sample are
from the $Hipparcos$ catalog (ESA 1997), while the radial velocities are from
the SIMBAD database.  The highlighted (filled) symbols 
show objects of special interest. Among the globular clusters,
$\omega$ Cen, NGC 362, and NGC 6779 are highlighted, as they are 
hypothesized to belong to the same parent galaxy (Section 2). 
One star in the NS97 sample, namely HD 106038,
 is particularly interesting, and it is highlighted in
Fig. 4 with a filled triangle. HD 106038 is overabundant in
Si ([Si/Fe] = 0.57), Ni ([Ni/Fe] = 0.18), and in s-process elements Y 
([Y/Fe = 0.49) and Ba ([Ba/Fe] = 0.49), when compared to
halo stars of similar metallicity (NS97, their Fig 4 and 5). 
At a metallicity of -1.26, HD 106038 displays
canonical $\alpha$ enhancement for metal-poor, halo stars produced
in a type II supernovae-dominated environment (NS97).
HD 106038's abundance patterns, in particular the enhanced s-process elements
are unusual among halo stars (NS97); however, they 
agree very well with the abundance patterns
of $\omega$ Cen stars shown by Smith {\it et al.} (2000).
HD 106038 is a main sequence star, with no radial-velocity evidence
of a companion likely to pollute the star with AGB ejecta (NS97).
\begin{figure}
\plotone{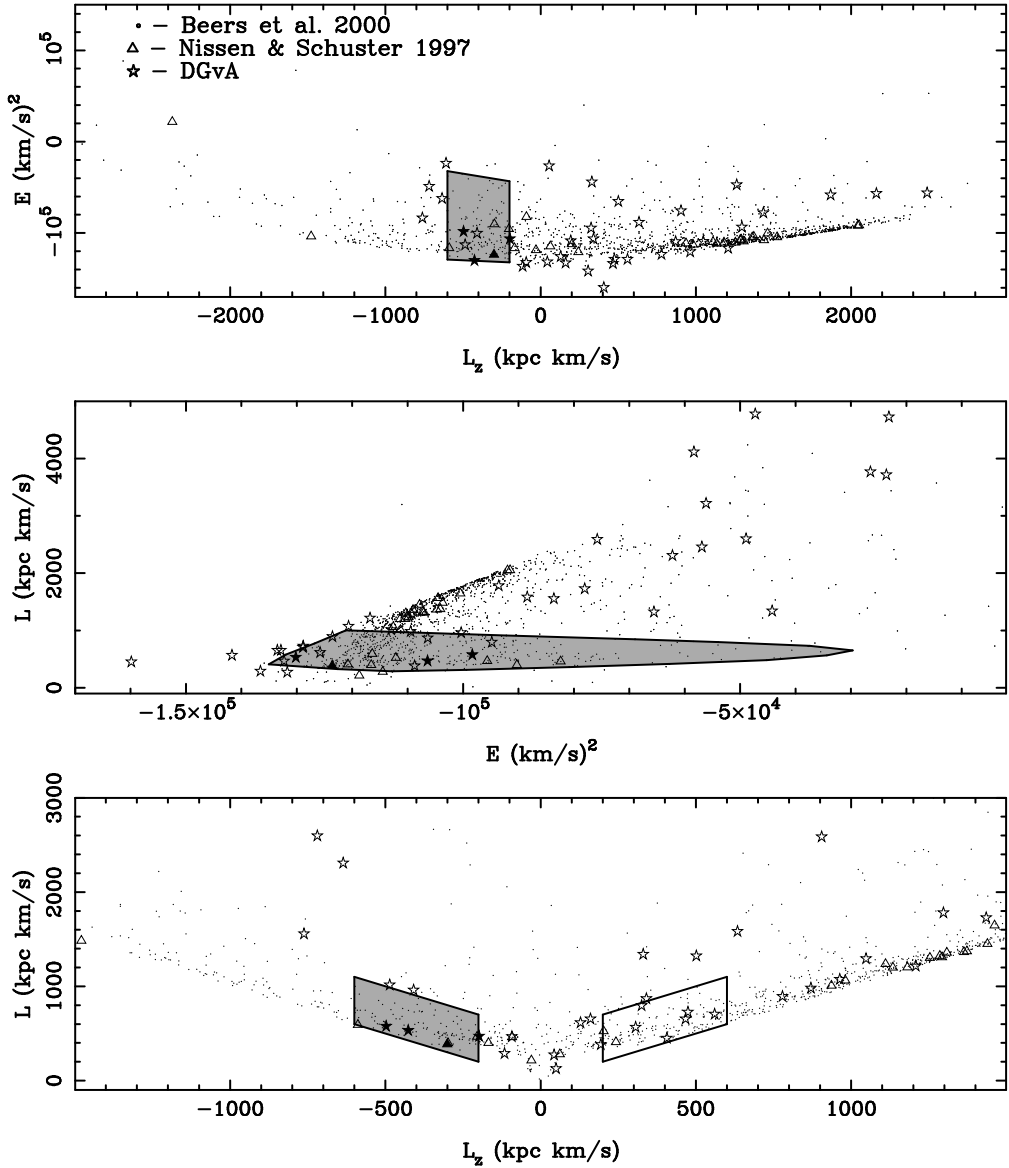}
\caption{Orbital parameters for Beers {\it et al.} (2000) and 
Nissen \& Schuster (1997) stars, and for globular clusters (DGvA).
Candidate stars from $\omega$ Cen's host galaxy are chosen to lie
in the shaded zone defined in the L-L$_{z}$ plot (bottom panel). 
The same area is also represented
in the E-L$_{z}$ plot (middle panel) and the L-E plot (top panel).
This area is defined by the three globular clusters NGC 362, $\omega$ Cen
and NGC 6779 (see text). In the L-L$_{z}$ plot, a similar zone in the
prograde domain is marked. This region has the same area, and is symmetrically 
located to the one in the retrograde domain.}
\end{figure}

In the plot of E as a function of L$_{z}$, the clusters $\omega$ Cen,
NGC 362, and NGC 6779 and HD 106038 lie in the retrograde region where the 
excess of stars was found in sample B of the B2000 catalog (Fig. 2, and 3).
The plots of L versus E, and L versus L$_{z}$ show that, indeed the
three clusters and HD 106038 lie in the same volume of the phase space.
Guided by the distribution of the three clusters, we define a box 
in the L-L$_{z}$ plot (shaded area in Fig. 4) 
aimed at selecting candidate objects in the NS97 and
B2000 data sets.  The corresponding area is overplotted in the
E-L$_{z}$ and L-E plots. 
It is interesting to note that, although the box
was defined by the L-L$_{z}$ pair to cover a restricted region, in
E, this region covers a large range of values.
Figure 5 shows the orbit inclination 
as a function of eccentricity for stars and clusters in the L-L$_{z}$
region defined above, and for stars in a symmetrically defined 
region at prograde orbits (shown by the contour in the bottom 
panel of Fig 4).
\begin{figure}
\plotone{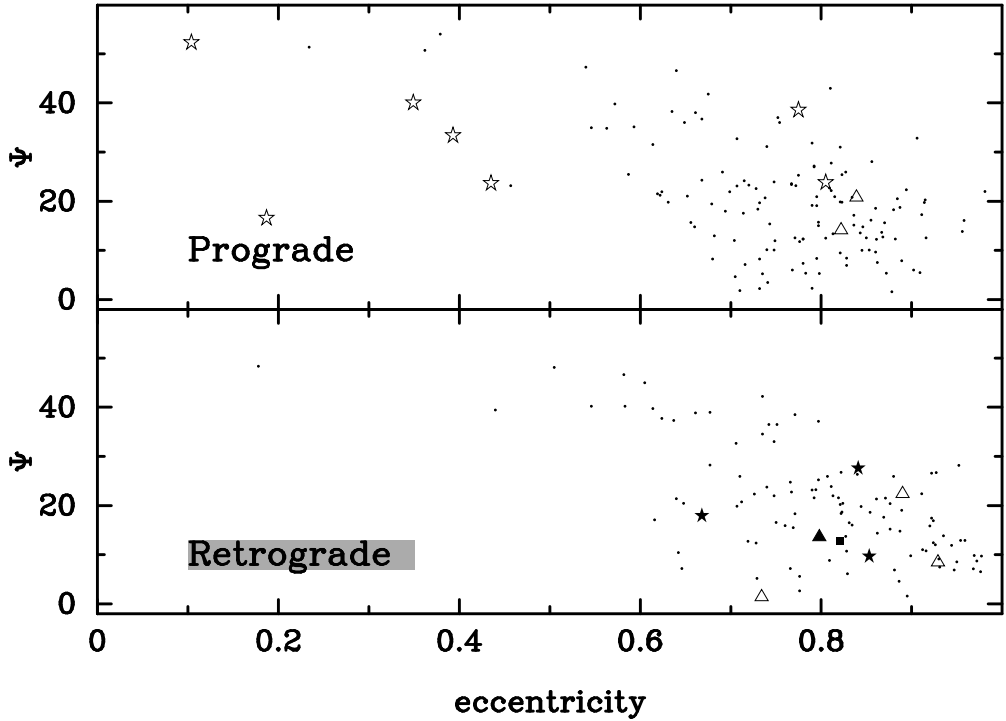}
\caption{Orbit inclination as a function of eccentricity
for the stars and clusters selected in the prograde and retrograde 
areas defined in Fig. 4. Symbols are as described in Fig. 4. The highlighted
(filled) symbols represent objects of particular interest, as follows.
The star symbols are the globular 
clusters $\omega$ Cen, NGC 362, and NGC 6779,
the triangle is HD 106038, and the square is V 716 Oph (see text).}
\end{figure}
There is not a significantly larger number of stars at $e > 0.8$ 
in the retrograde group (56 stars)
than in the prograde group (50 stars) 
in this latter plot. This is because we have included all stars from B2000.
However, selecting RR Lyrae with $e > 0.8$, we obtain
15 stars in the prograde box, and 29 in the retrograde box.
Also, in the retrograde domain, a structure
that starts from very high eccentricities ($ e \ge 0.9$) and very low 
inclinations ($\Psi \sim 8-10\deg$) toward $e \sim 0.8$ and $\Psi \sim 24\deg$
is apparent. This structure is also seen in the RR Lyrae population
(Fig. 3, bottom right panel). Although the region of $e > 0.8$ is well
populated in the prograde domain --- one globular cluster, and two
NS97 stars are found here --- there is no structure apparent
here, and there are very few stars at $e \ge 0.9$.

\section{RR Lyrae in Beers {\it et al.} Catalog}

We have seen that the excess population at L$z \sim -400$ kpc km/s,
$e \sim 0.85$, and $\Psi \sim 20\deg$ is enhanced when one considers
the RR Lyrae stars in B2000 (Section 4). 
In Figure 6 we show the periods as a function of metallicity
for RRab and RRc stars in B2000, and in $\omega$ Cen, for comparison.
The periods and the RR type were taken from the General Catalogue of
Variable Stars 4th edition (Kholopov {\it et al.} 1998, hereafter GCVS)
for the B2000 stars. For $\omega$ Cen, periods, RR type and
metallicities (based on the hk index) were taken 
from Rey {\it et al.} (2000).
The top panel shows the RRab (filled circles) and the RRc (open triangles)
stars in the prograde box defined in Fig. 4.
The middle panel shows those in the retrograde box, and the bottom panel
those in $\omega$ Cen. 
\begin{figure}
\plotone{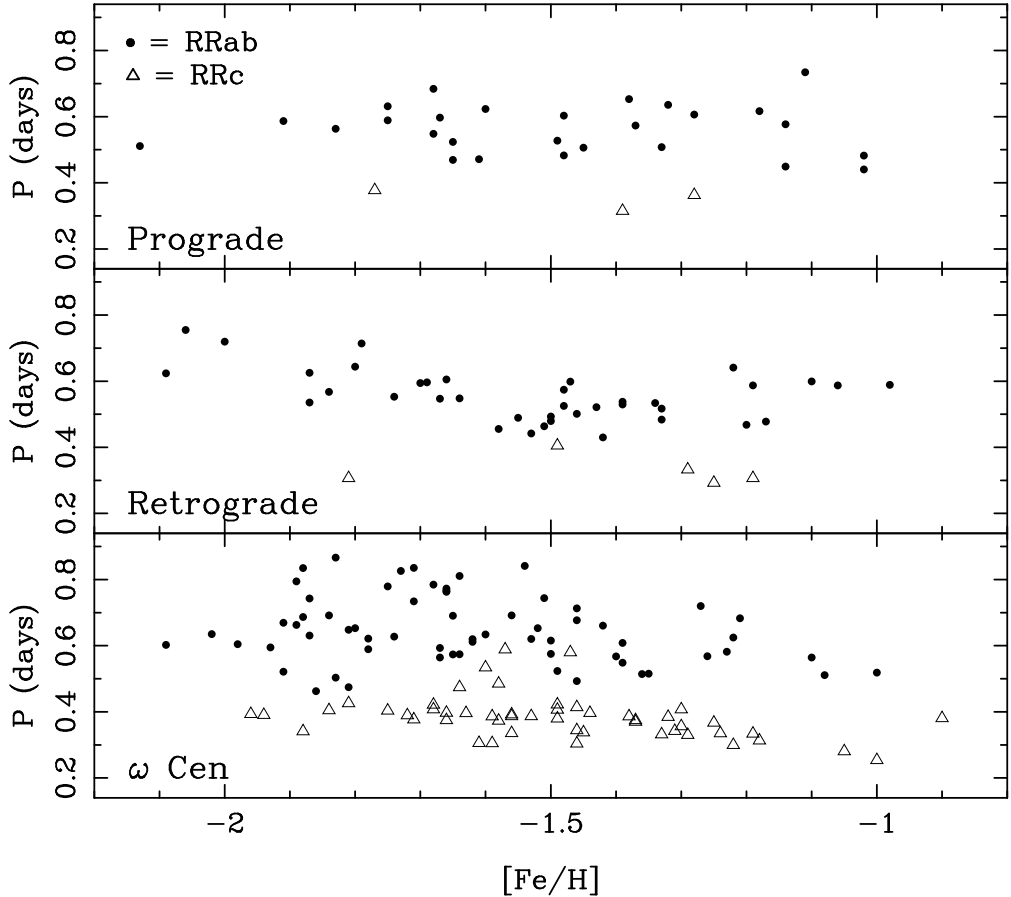}
\caption{The distribution of RR Lyrae periods with metallicities, for the 
prograde and retrograde regions defined in Fig. 4 (top and middle
panel respectively), and for stars in $\omega$ Cen (bottom panel).}
\end{figure}

The presence of RRab variables with P $ \ge 0.8$ days, and RRc
variables with P $ \ge 0.45$ days is characteristic of 
$\omega$ Cen (Fig. 6 with data from Rey {\it et al.} 2000,
see also, Clement \& Rowe 2000). These long period RR Lyrae stars
are seen in a larger abundance 
than in $\omega$ Cen only in two metal rich, bulge clusters
(NGC 6388 and NGC 6441, Pritzl {\it et al.} 2000).
There are no  such long period RR Lyrae stars in either the prograde
or the retrograde sample. 

The prograde sample shows a rather random distribution, with periods ---
on the mean --- lower than those in $\omega$ Cen. The retrograde sample
distinctly shows a high concentration of stars at P $\sim 0.5$ days and
[Fe/H] $\sim -1.5$, that is not seen in either the prograde sample or the
$\omega$ Cen sample.
It is predominantly these latter stars that comprise the excess of RR Lyrae
stars in the retrograde sample, when compared to the prograde
sample (Fig. 6 and Fig. 3, also Section 4).
The information from Fig. 6 is however insufficient to either support or
reject confidently the notion that part of the population of the RR Lyrae 
in the retrograde sample may resemble that in $\omega$ Cen.

Two stars that were classified as RR Lyrae in B2000, are in fact
W Virginis variables of short periods (P = 1.1 and 1.3 days), also known as 
BL Herculis variables, according to GCVS. 
This type of variable is found in globular 
clusters. $\omega$ Cen is the Galactic cluster with the 
largest population of BL Her objects: it has 5 such stars
according to Nemec, Linnell-Nemec, \& Lutz (1994). More recently, Kaluzny {\it 
et al.} (1997) have found three new BL Her candidates in $\omega$ Cen.
Interestingly, one of the two BL Her stars in B2000, namely V 716 Oph,
resides in the retrograde region defined in Section 4,
has an eccentricity $e = 0.82$, an inclination $\Psi = 13\deg$, and
a metallicity [Fe/H] = -1.55. V 716 Oph
is highlighted in the $e-\Psi$ plot of Fig. 5 with a filled square symbol.
The second BL Her star in B2000 is XX Vir. It has a metallicity
value [Fe/H] = -2.4, that is much lower than the metal-poor limit of
 stars in $\omega$ Cen.  

\section{A Simple Disruption Model}

The stars and globular clusters that have similar integrals of
motion as $\omega$ Cen, tend to have larger eccentricities than
$\omega$ Cen (Fig. 4, Section 4). Obviously, these candidates will have 
larger apocentric radii than $\omega$ Cen at a fixed pericenter radius,
in other words, they reside on orbits that are slightly more energetic than
$\omega$ Cen's. Under the hypothesis that these candidates belong to 
$\omega$ Cen or the disrupted system that once contained $\omega$ Cen, 
they ought to lie on trailing streams if they are on orbits slightly 
more energetic than  $\omega$ Cen's (e. g., Johnston 1998). 
We have used the tidal disruption model developed by Johnston (1998)
in order to see whether tidal debris from a system that now has 
the orbit of $\omega$ Cen can attain orbits such as those of
our candidates. We note that the Johnston (1998) model was developed for
the disruption of Galactic satellites that reside mostly in the
 outer Galaxy; therefore, quantitatively it may not be an accurate description
of the event. However, here we will present only a qualitative 
inspection. 

We start with a system of M = $5~10^6$ M$_{\odot}$ that 
has the orbit of $\omega$ Cen. We integrate back in time for 1 Gyr;
at each pericenter passage, the system is assumed to have lost
$30\%$ of it's mass.
In Figure 7 we show the spatial distribution
in the Galactic plane (left panel) and perpendicular to
 the Galactic plane (right panel) of the B2000 stars (dots), 
$\omega$ Cen's orbit (continuous line), and two trailing streams
(grey bands). The trailing streams
correspond to pericenter passage n$_{p}$ = -11 (light gray band; 918 Myr ago),
and to  pericenter passage n$_{p}$ = -10 (dark gray band; 835 Myr ago).
\begin{figure}
\plotone{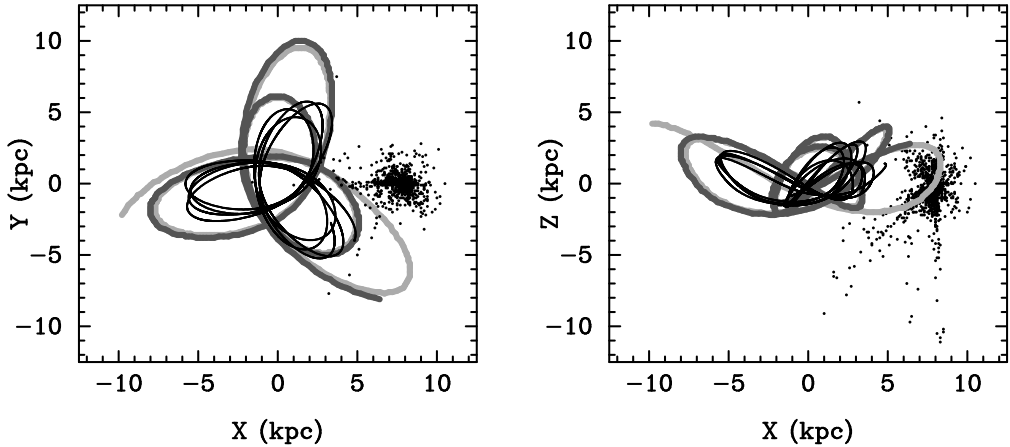}
\caption{The spatial distribution of the B2000 data sample (dots), 
$\omega$ Cen's orbit (thin dark line), and two trailing tidal streams:
pericenter passage = -11 (918 Myr ago; 0 is the pericenter passage 
 closest in time to present; light gray band), and pericenter passage = -10
(835 Myr ago; dark gray band).}
\end{figure}
The pericenter passage n$_{p}$ = 0 corresponds to the passage 
closest in time to present time. At n$_{p}$ = -11, the system had a total mass
of 3.6 10$^{8}$ M$_{\odot}$. The spatial location of the trails shows that
indeed, in the Solar neighborhood we can expect to find debris from this system.
These tidal streams are distributed along orbits that have the
shape of the orbits of our candidates:
high eccentricity and low orbital inclination.
We also note that all of the corresponding leading (lower energy)
tidal streams  are located in and close to the bulge.

A detailed N-body simulation of the disruption event has yet
 to be explored in order to better understand the process, and to be able to
quantitatively describe it. For instance, the mass of tidal debris
expected in the Solar neighborhood would be particularly useful 
in order to correlate with the excess of stars that we see in a
given domain of phase space (Section 4).
Similarly, the predicted kinematics of tidal debris, when compared
to that of candidate stars, can help demonstrate how viable, and
under what conditions, the accreted scenario for $\omega$ Cen is.

\section{Summary}

We have shown that a distinct population of stars 
with a metallicity range that inludes the mean metallicity of
$\omega$ Cen, and with $\omega$ Cen-like phase-space characteristics emerges
from the B2000 data. Choosing a metallicity
and orbital-parameter range (Section 4) such that we maximize
the ``signal'' of this population with respect to the ``noise'' of  
the halo, we obtain an excess population at 2.3-$\sigma$ level.
By considering the RR Lyrae stars in B2000, we also see
this population. We find that the excess RR Lyrae population is predominantly 
of RRab type, with periods of 0.5 days, and [Fe/H] $\sim -1.5$.

The candidates to have been torn from the system that once
contained/was $\omega$ Cen
have one main orbit property: they have a larger eccentricity 
($e \sim 0.8$) (i. e. orbital energy) than that of  $\omega$ Cen ($e = 0.67$).
Using the disruption model developed by Johnston (1998)
for the orbit of $\omega$ Cen, we find that trailing tidal debris  
 with orbit characteristics of those of the candidates are
found in the Solar neighborhood.

We also find that HD 106038, a single, main sequence
star with a chemical abundance pattern
very similar to that in $\omega$ Cen stars, in particular enhanced
s-process elements (NS97), has $\omega$ Cen-like orbital properties.
Similarly, V716 Oph in B2000 (a BL Her-type variable found in globular 
clusters of which $\omega$ Cen is most abundant) has $\omega$ Cen-like orbital 
properties, and a metallicity close to the mean metallicity of
$\omega$ Cen.

We identify two globular clusters as candidates for belonging
 to the system that
once contained/was $\omega$ Cen, NGC 362 and NGC 6779. The more metal
rich cluster, NGC 362 shows a deficiency in [Cu/Fe]
when compared to globular clusters of similar metallicity,
a deficiency seen so far only in $\omega$ Cen stars.

{\bf Acknowledgments}.
I am grateful to  M\'{a}rcio Catelan for his suggestions 
regarding the RR Lyrae stars, and to both M\'{a}rcio Catelan and
Terry Girard for numerous helpful discussions concerning this work.
This research has made use of the SIMBAD database, 
operated at CDS, Strasbourg, France.

\end{document}